\documentclass[pre,onecolumn]{revtex4}
\usepackage{graphicx}
\begin{document}

\title{GRENOUILLE - Practical issues: a quick manual}

\author{A. Christian Silva}

\affiliation{Department of Physics and Institute for Research in Electronics and Applied Physics, University of Maryland College Park, College Park, Maryland  20742}
\email{silvaac@physics.umd.edu}
\homepage{http://www.wam.umd.edu/~silvaac}
\date{\today}

\begin{abstract}
This paper is the result of setting up GRENOUILLE in the Nonlinear
Dynamics Laboratory at the University of Maryland at College Park.
With the experience acquired in the process of setting up
GRENOUILLE, this manual was compiled from literature and from
hand-on experience to serve as a quick guide, a step-by-step help
to construct GRENOUILLE and to understand some of its basic
principles.
\end{abstract}
 \maketitle

\section{Initial Considerations}

Frequency resolved optical gating (FROG) consists of an autocorrelator with its output going into a spectrometer connected to a CCD camera \cite{fpr,firstp,fg,rev}. The CCD camera records a spectrogram called the FROG trace (Fig.\ref{dt} - 3 dimensional graph with intensity as a function of time delay on the horizontal axis and wavelength on the vertical axis). From this spectrogram, the full electric field (intensity and phase) can be reconstructed with the aid of a numerical interactive algorithm. The mathematical form of the FROG trace depends on the nonlinearity used to generate it. For a second harmonic generating (SHG) crystal, the FROG trace is proportional to Equation (\ref{SHGF}), where the proportionality constant is divided out when the results are presented.

\begin{equation}
I_{FROG}(\omega,\tau)=\vert \int_{-\infty }^{\infty
} dt E(t)E(t-\tau) exp(i\omega t)\vert^{2}
\label{SHGF}
\end{equation}

GRENOUILLE (grading-eliminated no-nonsense observation of ultrafast incident laser light e-fields) is the simplest SHG FROG device ever built \cite{pat}.
A typical FROG device has a delay line, the nonlinear medium and a spectrometer. GRENOUILLE replaces the delay line by a Fresnel biprism and combines the spectrometer and the nonlinear medium into a thick SHG crystal.
The thick nonlinear crystal works as a frequency filter due to the large group velocity mismatch (GVM $\equiv 1/v_{g} ( \lambda _{0}/2 ) - 1/v_{g} ( \lambda _{0} )$) \cite{pat}. The working condition for the GRENOUILLE requires that $GVM\times L\gg \tau _{t}$, where L is the confocal parameter (the length of the crystal can be used for order of magnitude calculations) and $\tau _{t}$ the time duration of the pulse, that can be taken to be the full width at half maximum of the pulse being measured. Group velocity dispersion (GVD $\equiv 1/v_{g} ( \lambda _{0} - \delta \lambda /2) - 1/v_{g} ( \lambda _{0} + \delta \lambda /2)$) can be avoided by imposing the extra relation  $GVD\times L\ll \tau _{c}$, where $\tau _{c}$ is the pulse coherence time (approximately the reciprocal bandwidth, 1/$\Delta f$) \cite{pat}. These two conditions are combined in Equations (\ref{constr}) and (\ref{essent}), where TBP stands for time bandwidth product.

\begin{equation}
GVD\left( \frac{\tau _{t}}{\tau _{c}}\right) \ll \frac{\tau _{t}}{L}\ll GVM
\label{constr}
\end{equation}

\begin{equation}
\frac{GVM}{GVD}\gg TBP=\left( \frac{\tau _{t}}{\tau _{c}}\right)
\label{essent}
\end{equation}

Consider a transform limited (TBP $\sim 1$) input pulse with center wavelength of $800$ $nm$ and a FWHM of $\delta\lambda$ = $10$ $nm$. Consider also a $5$ $mm$ thick BBO crystal.  Equation (\ref{constr}) results in inequality (\ref{trnslim}) which shows that the crystal is appropriate for measuring the input beam \cite{pat}.

\begin{equation}
GVD = 20 fs/cm \ll 200 fs/cm \ll GVM = 2000 fs/nm
\label{trnslim}
\end{equation}

As a counter example, consider a pulse with TBP $\sim 10$.  Suppose that we have a bandwidth of the order of $\delta\lambda\sim 20$ $nm$ and hence a pulse time length of the order of $100$ $fs$. Equation  (\ref{constr}) is not satisfied for the $5$ $mm$ BBO crystal. Table (1) presents a comparison for the case TBP $\sim 10$ and center wavelength $\lambda_{c}\sim 800$ $nm$.

\begin {center}
\begin{tabular}{|c|c|c|c|c|}
\hline
$\delta \lambda (nm)$ & $TBP\times GVD(fs/cm)$ & $\tau _{p}/L(fs/cm)$ & $GVM(fs/cm)$& Works? \\ \hline
20 & 330 & 200 & 2000 & No \\ \hline
10 & 200 & 400 & 2000 & No \\  \hline
5 & 113 & 800 & 2000 & Might\\
\hline
\end{tabular}
\end {center}

\section{Experimental Setup}

\begin{figure}[t]
\scalebox{0.8}{\includegraphics{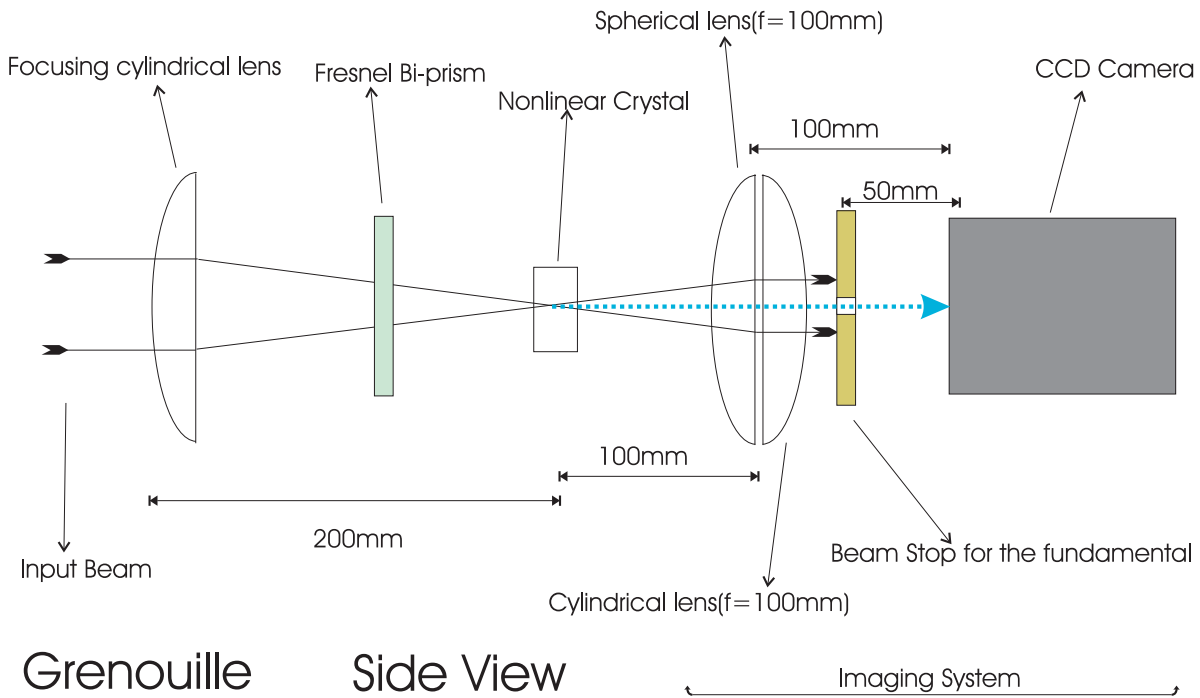}}
\caption{Grenouille's experimental setup from a side view. The
picture is not drawn to scale. Nonlinear crystal: BBO dimensions
$5\times 10 \times 10$ $mm$. Fresnel biprism has an apex angle of
$168^{0}$ and almost square base of $ 1 \times 1$ $in$. The camera
used is a Pulnix TM-72EX with a filter for the fundamental beam
adapted to its opening.} \label{sync}
\end{figure}

GRENOUILLE simplest experimental setup is presented in Figure \ref{sync}, where the vertical axis of a measured FROG trace is automatically aligned to be wavelength and the horizontal axis, time delay. Folding mirrors can be included to further reduce the final size of the setup (the commercial GRENOUILLE is a small rectangular box of $5 cm \times 15cm \times 25cm$)but that does not change the experimental set in any essential way.

The alignment has to closely achieve Figure \ref{sync} configuration with the second harmonic going in the center of the CCD camera and the fundamental beams being blocked by the slit. All the distances shown in Figure \ref{sync} are within $2$ $mm$ for an ideal FROG trace (It seems enough to use a ruler to measure these distances). It is "easy" to get some SHG "trace", the problem is to have a good quality trace and also to be sure what has been measured is indeed a FROG trace. After the initial alignment, fine adjustments to the imaging system as well as the position of the input cylindrical lens have to be done. The distance between the input cylindrical lens and the crystal is found by maximizing the intensity of the second harmonic beam after the crystal. The position as well as function of each element in GRENOUILLE is now discussed.

\subsection{Input beam}

The size of the input beam determines the range of wavelength and time delay seen in the FROG trace and consequently limits the spectral range as well as time range of the pulse to be measured.

The spectrometer in traditional FROG apparatus is replaced by the thick non-linear crystal and by the size of the input beam. The thick crystal determines the resolution and the input beam the range.

\begin{figure}[t]
\scalebox{0.8}{\includegraphics{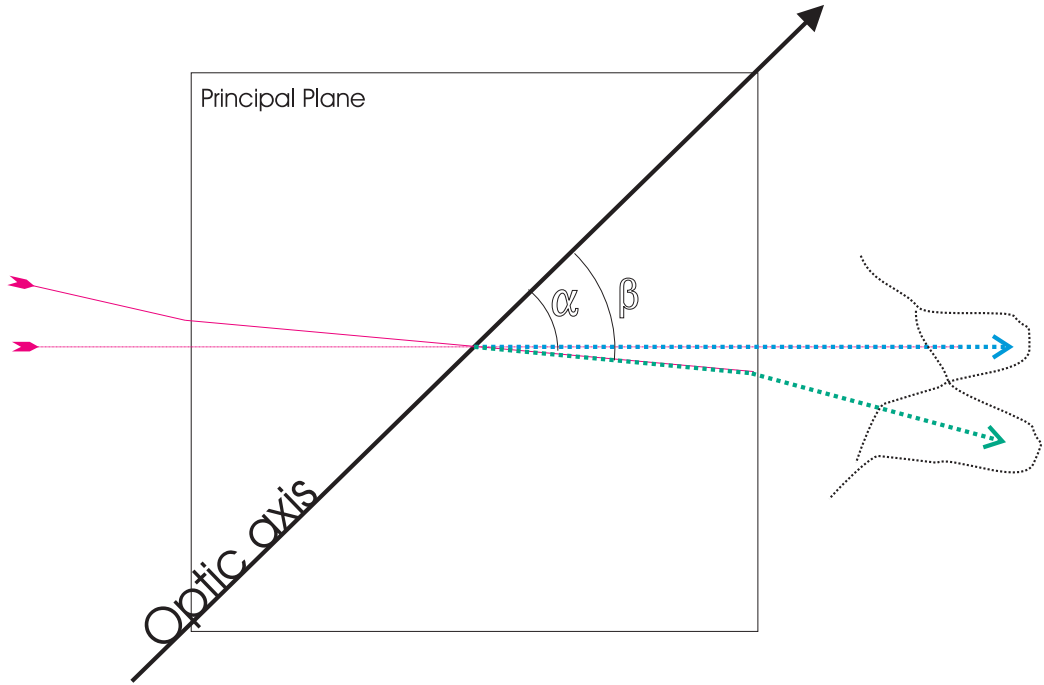}}
\caption{Neglecting the polarization, this shows that although the
crystal was cut for a certain wavelength (say 800nm, the center
beam with angle $\alpha$)the beam coming with an angle $\beta$,
close to $\alpha$, will also phase match as long as the pulse has
a wavelength component that phase matches at $\beta$. If $\beta$
is the most extreme angle in the range of angles of the input beam
inside the crystal that phase matches, the wavelength range of the
setup is the wavelength that frequency doubles between $\alpha$
and $\beta$. For this setup the polarization is perpendicular to
the surface of the figure for the incoming beam and on the plane
of the figure for the SHG beams (blue and green beams in figure).}
\label{inpb}
\end{figure}

The range of angles covered by focusing the input beam inside the nonlinear crystal determines the wavelength range that can be measured by GRENOUILLE. Figure \ref{inpb} shows that only a small wavelength range is phase matched at each angle (the smallness of wavelength range is determined by the crystal filtering). This mechanism creates at each angle $\alpha$ or $\beta$ a small range of wavelength that is further mapped into the CCD camera.

Consider the GRENOUILLE constructed with a BBO crystal and designed to work with input beams from $10$ $mm$ to $20$ $mm$ at a center wavelength of $810$ $nm$. Beams with waist of about $10$ $mm$ have a range of angles inside a BBO crystal of about $1.72^{0}$. Phase matching angle for $810$ $nm$ is about $28.9 ^{0}$, for $860$ $nm$ is about $27.3 ^{0}$ \cite{hand,thick}. From these values, the wavelength range with a $10$ $mm$ input beam is about $810 \pm 25$ $nm$.

The wavelength range increase of the GRENOUILLE can not be achieved by simply increasing the vertical size of the input beam. The confocal parameter has to be considered, since it is the effective length of the crystal. Bigger input beam waist also implies smaller confocal parameter. Smaller confocal parameter might also imply violation of Expression (\ref{constr}) , the fundamental condition for GRENOUILLE to work.

\begin{figure}
\scalebox{0.8}{\includegraphics{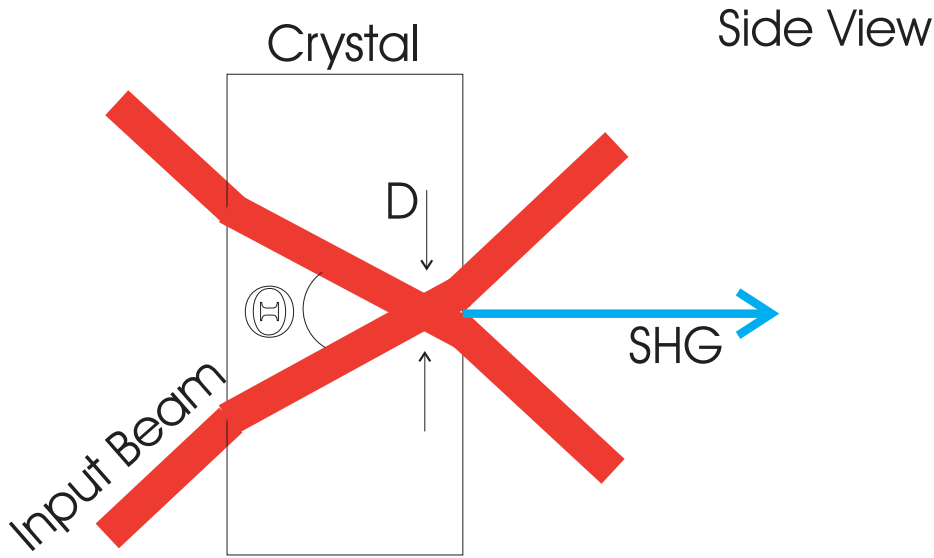}}
\caption{The effect of the input cylindrical lens and the Fresnel biprism is to create two crossing strips of the input beam inside the crystal. The width of these strips is responsible for the time delay in the FROG trace (blue beam).}
\label{delt}
\end{figure}

The delay line in traditional FROG traces is replaced by the horizontal beam waist and the Fresnel biprism. The range of the time delay axis is given by the size of the horizontal beam waist crossed inside the crystal through the action of the biprism (Fig. \ref{delt}). The expression for the delay is just given by the size of the horizontal crossed trace inside the crystal ($D$) and by the angle ($\theta$) formed by the crossed beams (Eq. \ref{dely} ) \cite{thick,practic}.

\begin{equation}
\Delta \tau =\frac{D\times \sin \theta }{c}
\label{dely}
\end{equation}

For an input beam size of $10$ $mm$, BBO crystal, and a Fresnel biprism with apex angle $168^{0}$, the delay axis is $\Delta\tau \leq 2.3$ $ps$.

\subsection{Fresnel Biprism}

The function of the Fresnel biprism \cite{pat} is to create from the input beam two beams that cross inside the nonlinear crystal (Fig. \ref{delt}). The position of the biprism between the focusing lens and the nonlinear crystal can be chosen to adjust the range of the delay axis. The maximum delay axis for a given crystal is calculated with Equation \ref{dely} where the input horizontal beam size plays a fundamental role discussed in the previous section.

The optimal size of the delay axis does not need to be the maximum delay size achievable with a given GRENOUILLE setup. The required condition for the range of the delay axis is to include the majority of the FROG trace. The FROG trace has to show all its intensities including its outer edges (horizontal outer edges may be clipped for time delay axis that are not wide enough). The smallest delay axis range that shows the full FROG trace is the optimum delay axis range. Smaller delay axis give better quality FROG traces since more of the fundamental beam is concentrated on the area of the second harmonic conversion increasing the intensity of the trace.

For an input beam of $10$ $mm$ (horizontal waist), the distance between the BBO crystal and the Fresnel biprism that maximizes the delay axis is about $100$ $mm$. In practice, one can search for the best position by looking at the FROG trace while moving the Fresnel biprism in between the crystal and the focusing lens.

The Fresnel biprism can be mounted on a fixed optical mount. Rotation on the perpendicular plane of figure \ref{sync} or tilt towards the nonlinear crystal or the focusing cylindrical lens is not needed, as long as the input focusing cylindrical lens can be rotated on the plane perpendicular to the plane of figure \ref{sync}. The rotation degree of freedom is critical for the fine adjustments on GRENOUILLE's alignment. Rotating the input cylindrical lens, helps aligning the reflection symmetry about the wavelength axis inherent in the GRENOUILLE FROG traces (SHG FROG trace is even with respect to the delay).

\subsection{Nonlinear Crystal}

The Nonlinear crystal is chosen to meet the requirements given in Equations (\ref{constr})  and (\ref{essent}).  The combination of the input beam and nonlinear crystal is equivalent to a spectrometer where the resolution is determined by the nonlinear crystal.

The second harmonic generated pulse intensity from a nonlinear crystal is given in the frequency space ($\omega$) by Equation (\ref{spec}), where L is the length of the crystal, $\Delta$K the phase matching condition and GVM the group velocity mismatch between the fundamental and the SHG beam \cite{diels}. The $sinc^{2}$ ($sinc = sinx/x$) term limits the bandwidth for the SHG pulse. The bandwidth becomes narrower with the increase of GVM and also narrower with the length ($L$) increase of the crystal. Hence to increase filtering, maximum GVM$\times L$ is a desired quality for the Nonlinear crystal, which leads to the fundamental conditions for GRENOUILLE to work (Eqs. (\ref{constr},\ref{essent})).

\begin{equation}
S(\omega )\propto L^{2}\times sinc^{2}\left[ \left( GVM\times \omega
-\Delta k\right) \frac{L}{2}\right] \times \left| \int_{\infty }^{\infty
}\xi \left( \omega -\omega ^{^{\prime }}\right) \xi \left( \omega ^{^{\prime
}}\right) d\omega ^{^{\prime }}\right| ^{2}
\label{spec}
\end{equation}

The bandwidth of the filtering process is calculated using Sellmeier equations \cite{hand,thick} and depends on the confocal parameter of the focused input beam inside the nonlinear crystal. Most of the SHG takes place along the confocal parameter which then defines the effective length of the crystal. The minimum bandwidth in a setup for a given nonlinear crystal is achieved with a confocal parameter of the length of the crystal. A BBO crystal $5$ $mm$ long with the input beam at $800$ $nm$ has the minimum bandwidth of about $2.5$ $nm$. The bandwidth for a confocal parameter of $2$ $mm$ is approximately $2.8$ $nm$ even though the crystal is still $5$ $mm$ long \cite{pat}.

The most important point in aligning the nonlinear crystal is to set the principal plane on the vertical plane of the table (for the set up of Figs. \ref{sync} and \ref{inpb}). The polarization of the incoming beam should be parallel to the line focus of the incoming cylindrical lens. This will guarantee that the vertical axis is the wavelength axis. It is important to phrase this point since one can always have SHG generation, even with the crystal at the wrong orientation. With the crystal correctly set, one should see 3 SHG beams at the same height. The FROG trace is the center beam. Translating the crystal on the plane perpendicular to the incoming beam might be helpful to avoid defects on the surface of the crystal. Tilting and rotating the crystal helps on the alignment since phase matching is angle dependent.

\subsection{Imaging System}

The function of the imaging system is to take the FROG trace generated by the nonlinear crystal and image it on the surface of the CCD array. The combination spherical lens and cylindrical lens (Fig. \ref{sync}) is assembled back to back on a mount with rotational degree of freedom. These combination creates a vertical focus of $f = 100$ $mm$ and a horizontal focus of $f = 50$ $mm$.

\begin{figure}[b]
\scalebox{0.8}{\includegraphics{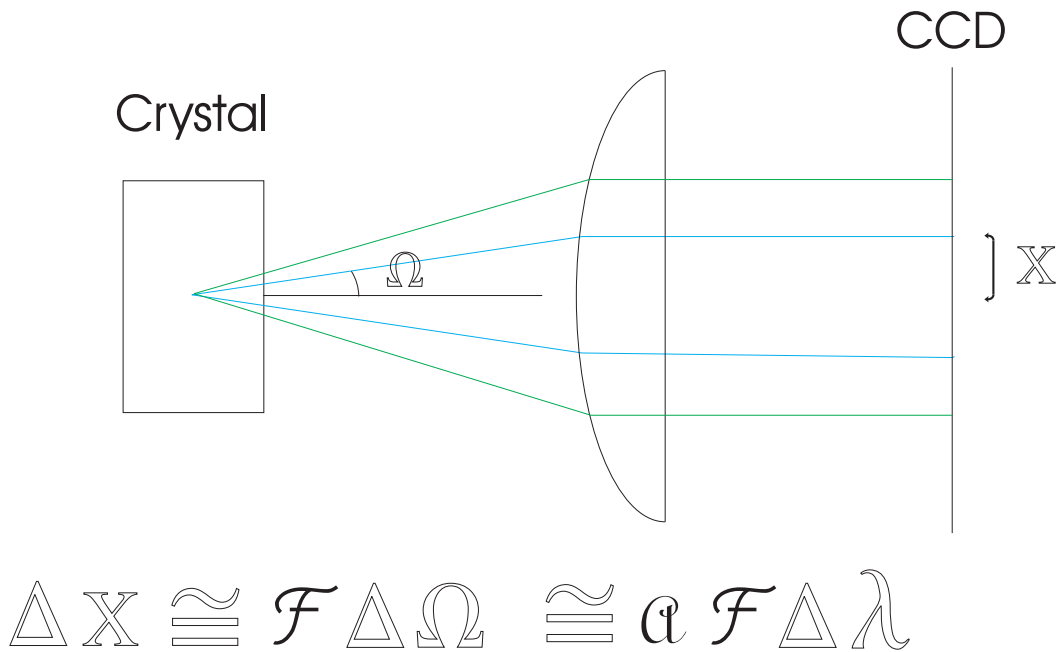}}
\caption{Side view of schematic representation of the equivalent lens on vertical axis of the imaging system for the setup of figure \ref{sync}. The represented lens is cylindrical and has a focus $f$=$100$ $mm$. The SHG beam coming from the nonlinear crystal has wavelength as a function of angle $\lambda (\Omega)$. The imaging lens takes $\Delta\ x = Const^{0}\times\Delta\Omega = Const\times\Delta\lambda$. Knowing the center wavelength $\lambda (\Omega = 0)$ from an independent measurement, the wavelength along the $x$ axis can be found. Note that the linear relations between $x$ and $\lambda$ are only valid for small angles ($\Omega$).}
\label{img}
\end{figure}

The vertical direction of the imaging lens maps the wavelength dependent angle from the FROG trace onto the surface of the CCD array as in Figure \ref{img}. The position on the camera ($x$) is a nearly linear relation of the wavelength ($\lambda$).

The time delay axis is collected by the horizontal direction of the imaging lenses and directed into the CCD camera. In this set up (Fig.\ref{sync} ), the size of the FROG trace along the time delay axis on the CCD array is equal to the one produced by the crystal.

The position of the CCD camera is set to have the FROG trace in the center. The FROG algorithm needs as an input parameter the center wavelength, which is half of the center wavelength of the pulse being measured only at the center of FROG trace.

The resolution of the Frame grabber and CCD camera has to be higher if the FROG trace is more complex. For simple traces (such as Gaussian pulses), having a FROG trace of size $64\times64$ pixels is enough to recover the correct phase and intensity of the input pulse \cite{practic}. For this setup (Fig.(\ref{img})), the resolution used is $512\times480$ pixels.

The overall alignment of the GRENOUILLE has to achieve a time delay symmetrical FROG trace located at the center of the frame grabber view area.

\subsection{Calibration}

\begin{figure}
\scalebox{0.8}{\includegraphics{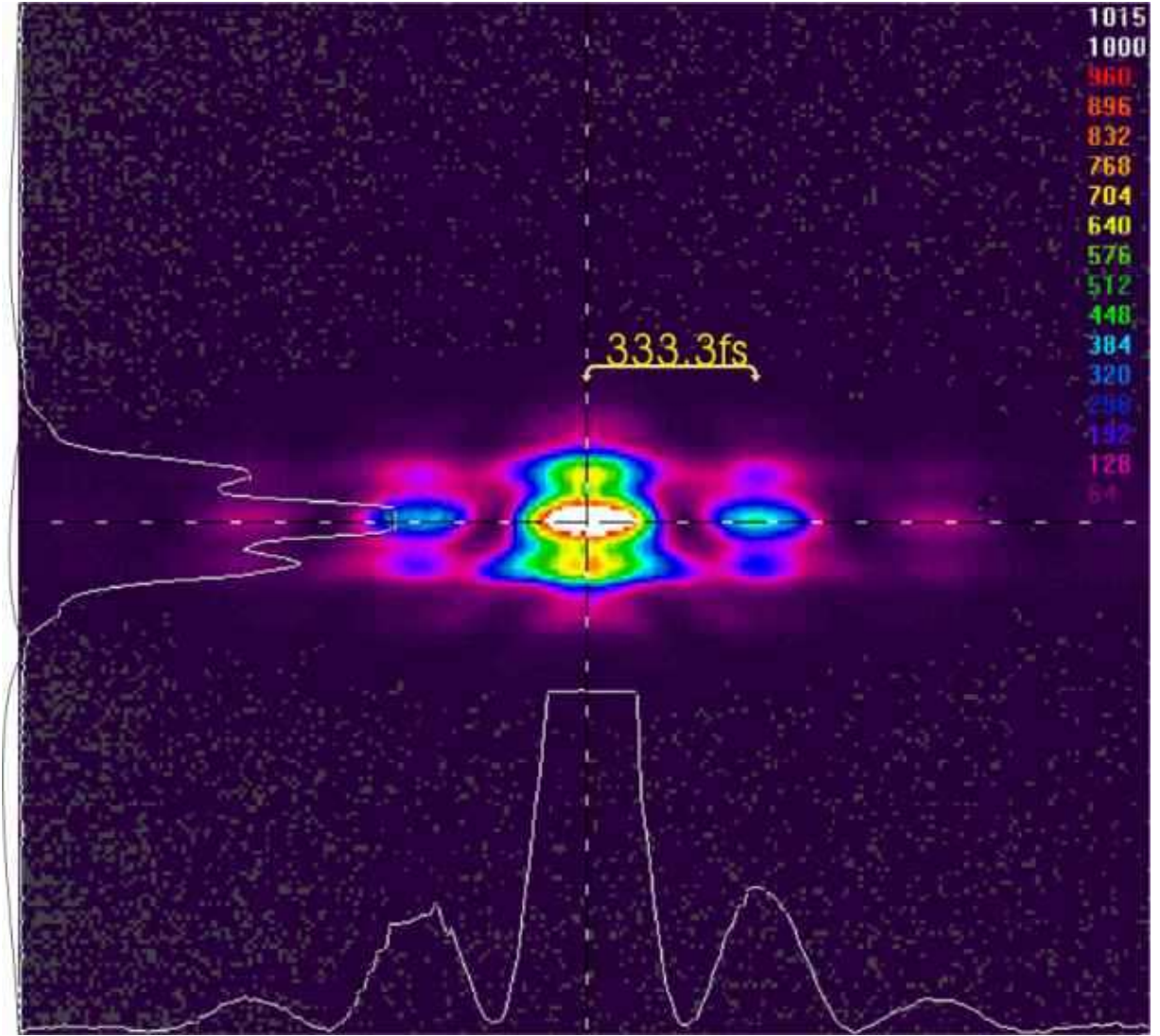}} \caption{FROG trace from
a train of pulses separated by multiples of $333.33$ $fs$. Such
FROG trace is produced by an air spaced etalon in the path of a
transform limited $810$ $nm$ Spectra-Physics Tsumami Ti:sapphire
oscillator with pulse width of $\sim 190$ $fs$. The horizontal
axis is time delay and the vertical axis is wavelength.}
\label{dt}
\end{figure}

The scale for the wavelength axis and time delay axis is found with the aide of an Etalon. The etalon also helps in lifting the time direction ambiguity inherent in GRENOUILLE (SHG) FROG \cite{fg}.
An air spaced etalon with spacing $d$ produces pulses separated by $2d/c$ in time, where $c$ is the velocity of light in air. The resultant pulse trace has features that can be used to calibrate both time and wavelength axis. The direction of time is also known since the second pulse coming out of the etalon is always of lower intensity than the first.
Figure \ref{dt} shows a FROG trace after an air spaced etalon with $d = 50$ $\mu m$.
The features used to calibrate GRENOUILLE are the modulations both in the horizontal and vertical axis. The modulations on the delay axis are separated by $2d/c$. Measuring the number of pixels separating the modulations gives the scale for the delay axis ($2d/(c\times \#pixels)$).

The vertical modulations are related to the horizontal modulations by $\Delta \tau \Delta f=1$. Using this relation to find the wavelength separation of the vertical modulations, one only needs to divide the wavelength separation by the number of pixels.

Typical values for the set up of Figure \ref{sync} and for the trace in Figure \ref{dt} are, for the horizontal axis (time delay), $8$ $fs/pixel$ and for the vertical axis (wavelength) $0.14$ $nm/pixels$.

Errors in the calibration induce errors in the recovered pulse. The size of the error in the recovered electric field is at least of the same size of the error in the calibration. An error of 20\% in the time delay axis leads to an error of at least 20\% in the time width of the recovered electric field. The exact influence of the error in the calibration depends on the complexity of the pulse to be measured and generally is bigger for more complicated pulses \cite{shp}.

\section{FROG Pulse Retrieval Algorithm}

The final goal of FROG is to determine the complex electric field from the FROG trace. This is done with the aid of a numerical algorithm that starts with an initial guess of the complex electric field and iterates to the measured electric field. The algorithm assumes two constraints. The first is the form of the electric field generated by the nonlinearity in use. For SHG, the electric field is given by Equation (\ref{elem}). The second constraint is the experimentally acquired FROG trace which is given by Eq.(\ref{elen}) or Eq.(\ref{SHGF}). Extra constrains are reported to create instabilities in the algorithm (for instance overflow), including adding an independent measured spectrum as one of the constrains \cite{rev}.

\begin{equation}
E_{sig}(t,\tau)= E(t)E(t-\tau)
\label{elem}
\end{equation}

\begin{equation}
I_{FROG}(\omega,\tau)=\vert \int_{-\infty }^{\infty
} dt E_{sig}(t,\tau) exp(i\omega t)\vert^{2}
\label{elen}
\end{equation}

\begin{figure}
\scalebox{0.7}{\includegraphics{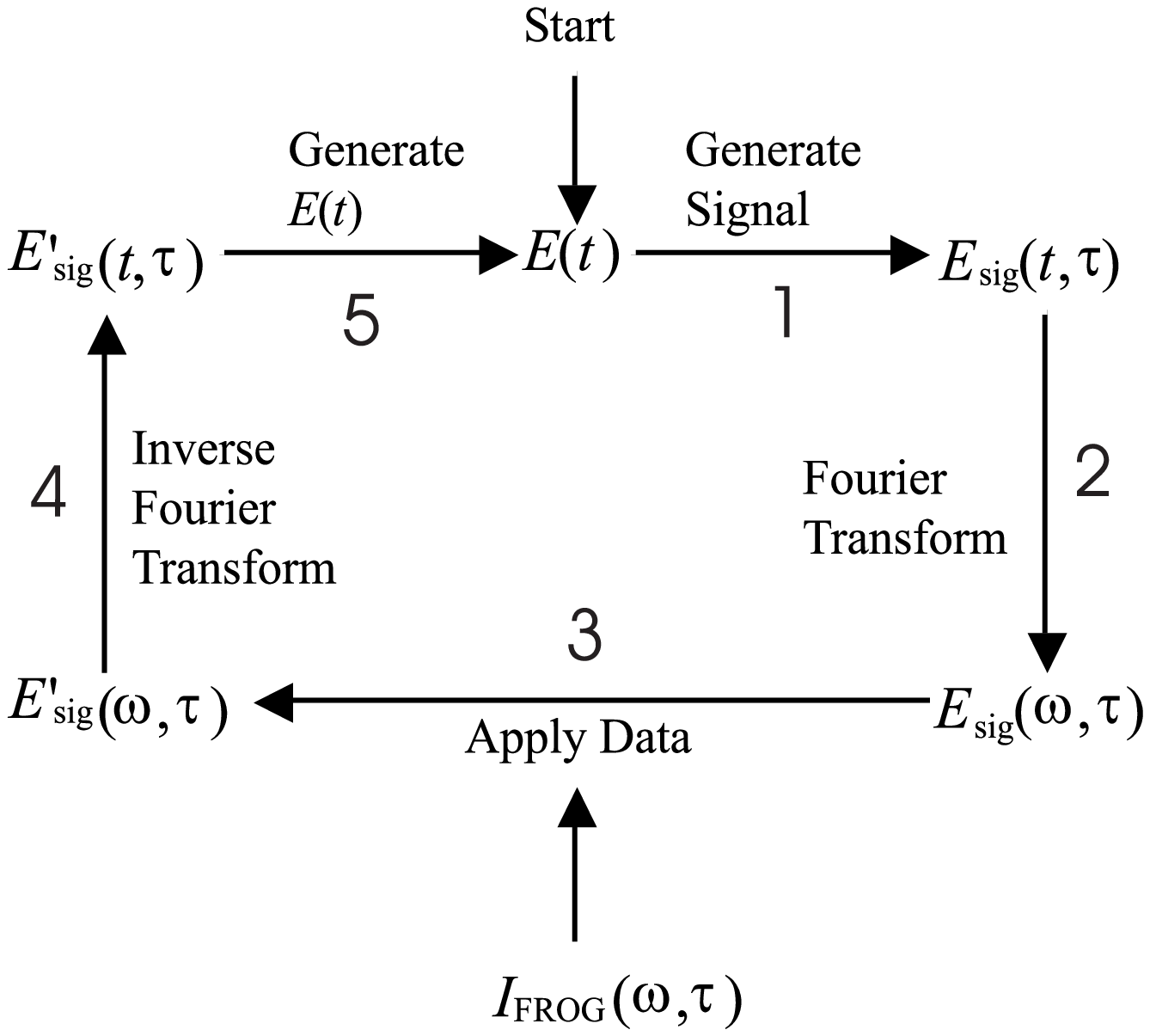}}
\caption{General diagram for the FROG algorithm adapted from \cite{fg} . For SHG FROG and GRENOUILLE the signal generation is given in Eq.(\ref{elem}).The loop 1-5 continues until desired error level is reached.}
\label{aw}
\end{figure}

The algorithm works as presented in Figure (\ref{aw}). The initial guess can be any electric field, the outcome should not depend on the initial guess, but the convergence time is minimized for clever guesses. The commercial code (Femtosoft) gives 4 options for initial guesses but does not allow for external input.

Once the initial guess is chosen the algorithm enters into a loop. The steps in the loop depend on the details of the algorithm. There are at least 8 different algorithms that can be used \cite{fg,gpr}. The most common can be found in the commercial versions of the FROG algorithms (Femtosoft and MarkTech), they are: the Basic FROG (or Vanilla FROG, first algorithm created), Generalized Projections, Short-cut Generalized Projections,Projections Over-Step, Intensity Constrained Basic FROG, Over-Correction Basic FROG and Multidimensional Minimization Technique \cite{fg}.

The algorithm that has been reported as most robust and responsible for the success of FROG is Generalized Projections \cite{gp}. Generalized Projection was shown to converge to the real pulse for all FROG traces where the other algorithms fail. Generalized Projections works as follows. The first step (step 1), $E_{sig}$ is constructed with Equation (\ref{elem}) and then $I_{FROG}$ is calculated with a fast Fourier transform (step 2). In step 3 the calculated $I_{FROG}$ has its magnitude replaced by the magnitude of the measured FROG trace (Eq.(\ref{rep})). In step 4, $E_{sig}^{'}(t,\tau)$ is constructed by taking the inverse Fourier transform. Until now all the steps are also true for the vanilla FROG algorithm. While the Vanilla FROG takes step 5 only by integrating $E_{sig}(t,\tau)$ on $\tau$ to get back $E(t)$, Generalized Projections takes step 5 by introducing distance minimization ($Z$) between $E_{sig}^{'}(t,\tau)$ and $E(t)E(t-\tau)$ to get $E(t)$ back (Eq.(\ref{min})). Once the Step 5 is complete, the cycle is reinitiated.

\begin{equation}
E_{sig}^{'}(t,\tau)=\frac{E_{sig}(t,\tau)}{\vert E_{sig}(t,\tau) \vert} \times \lbrack I_{FROG}(\omega,\tau) \rbrack ^{1/2}
\label{rep}
\end{equation}

\begin{equation}
Z=\sum_{t,\tau =1} ^{N} \vert E_{sig}^{'}(t,\tau) - E(t)E(t-\tau)\vert^{2}
\label{min}
\end{equation}

\begin{equation}
G=\lbrack \frac{1}{N^{2}}\sum_{\omega,\tau =1} ^{N}  \lbrack I_{FROG}(\omega,\tau) -\vert E_{sig}(\omega,\tau)\vert^{2} \rbrack ^{2} \rbrack ^{1/2}
\label{err}
\end{equation}

The error in the pulse retrieval ($G$) is measured with the FROG error (Eq.(\ref{err})), where $N$ is the grid size of the FROG trace and $\vert E_{sig}(\omega,\tau)\vert^{2}$ is normalized to a pick value of unity before $G$ is calculated \cite{shp}. The size of the FROG error decreases after each loop and typical error values depend on the quality of the trace and on the size of the grid. The program converges to the correct pulse for values of the error less than $10^{-4}$ for noise-free pulses (theoretical traces). It is not uncommon to achieve errors less than $0.5\%$ for good quality simple SHG traces $128\times128$ in size.

The quality of the traces and hence the minimum error ($G$) level achievable when using the FROG algorithm depends on the noise level when recording the trace. For multiplicative noise ($I_{FROG-Noise}=I_{FROG}\times(1+noise)$), $G\sim e\times(TBP/N)^{1/2}$, where TBP is time-bandwidth product of the pulse, $e$ is the error in the trace data points where the trace is nonzero and $N\times N$ is the array size. For additive noise ($I_{FROG-Noise}=I_{FROG}+ noise$) the error is $G\sim e$ \cite{fg,noise}.

Despite the success of Generalized Projections to find the intensity and phase of a light pulse, this technique, as well as any other technique, is not proven to converge for all pulses. The strategy adopted in the commercial programs, is to combine several algorithms into the program and to switch between them during a run. The commercial Femtosoft algorithm switches between strategy everytime the error does not decrease by $0.5\%$. This approach hopes to combine the strength of all the strategies and eventually cover all the possible FROG traces found in nature.

\subsection{Practical issues when using Femtosoft FROG pulse recovery program}

If pulse to be recovered is not theoretical, noise filtering is appropriate. It is customary to first remove the lowest pixel and then the edge of the data in the window after the trace has been imported into the program. This operation should be done only once to avoid clipping the low intensity data.

After noise filtering the data has to be extracted for the program to initiate its cycle. It is possible to extract only a small window around the FROG trace avoiding the background noise by selecting with the mouse the desired area. This will result into a smother recovered electric field but the algorithm has a recently discovered bug. This operation will rescale the wavelength axis incorrectly.

\begin{acknowledgments}

I thank Pat O'Shea and Mark Kimmel for the indispensable counsel
of how to set up GRENOUILLE. In special, I thank Pat O'Shea for
detailed discussion on the basics of FROG and GRENOUILLE. I also
thank Bhaskar Khubchandani with direct help in setting up
GRENOUILLE. Finally, I thank Prof. Rajarshi Roy for advice,
guidance and logistic support, without which this work would have
not been possible.

\end{acknowledgments}

\end{document}